\documentclass[iop,apj]{emulateapj}                
\usepackage{amsmath,amssymb,amstext}

\usepackage[breaklinks,colorlinks,citecolor=blue,linkcolor=magenta]{hyperref} 

\usepackage[all]{hypcap} 

\usepackage{natbib}
\usepackage{multirow}
\shorttitle{Constraining the nature of dark matter with star-formation histories of dwarf galaxies}
\shortauthors{Chau et al.}

\begin{document}

\title{Constraining the nature of dark matter with the star-formation history of the faintest Local Group dwarf galaxy satellites}
\author{Alice Chau$^1$, Lucio Mayer$^1$ and Fabio Governato$^2$}
\affil{$^1$Center for Theoretical Astrophysics and Cosmology, Institute for Computational Science, University of Zurich,
Winterthurerstrasse 190, CH-8057 Z\"urich, Switzerland \\
$^2$Astronomy Department, University of Washington, Box 351580, Seattle, WA 98195-1580}

\begin{abstract}
$\Lambda$-Warm-Dark Matter ($\Lambda$WDM), realized by collisionless particles of $1-3$~keV, has been proposed as alternative scenario to $\Lambda$-Cold-Dark Matter ($\Lambda$CDM) for the dwarf galaxy scale discrepancies. 
We present an approach to { test} the viability of such WDM models using star-formation histories (SFHs) of the dwarf
spheroidal galaxies (dSphs) in the Local Group. 
We compare their high-time-resolution SFHs
with the collapse redshift of their dark halos in CDM and WDM.
Collapse redshift is inferred after determining the subhalo infall mass. This is based on the dwarf current mass inferred from stellar kinematics, combined with cosmological simulation results on subhalo evolution.
WDM subhalos close to the filtering mass scale, forming significantly later than CDM, are the most difficult to reconcile with early truncation of star-formation ($z\geq 3$).
{ The Ultra-Faint Dwarfs (UFDs) provide the most stringent constraints. Using six UFDs and eight classical dSphs, we show that a 1~keV particle is strongly disfavored, consistently with other reported methods. Excluding other models is only hinted for a few UFDs. Other UFDs for which the lack of robust constraints on halo mass prevents us carrying out our analysis rigorously, show a very early onset of star-formation
that will strengthen the constraints delivered by our method in the future.}
We discuss the various caveats, notably the low number of dwarfs 
with accurately determined SFHs and the uncertainties when determining 
the subhalo infall mass, most notably the baryonic physics.
Our preliminary analysis may serve as pathfinder for future investigations that will combine accurate SFHs for local dwarfs with direct analysis of WDM simulations with baryons. 

\end{abstract}

\keywords{galaxies: star-formation, dark matter}
\maketitle

\section{Introduction}
The nature of the dark matter still remains a mystery. Most of the observational evidence point toward cold dark matter: The $\Lambda$CDM cosmology model describes accurately the large-scale structure of the universe, reproduces naturally all the properties of the
Cosmic Microwave Background and produces a scenario for galaxy formation that is receiving increasing confirmations from various
observational diagnostics. Indeed, small scale problems that used to be vexing for two decades, 
such as the angular momentum problem in disk galaxies or the
shape of the rotation curves of gas-rich dwarf galaxies, are largely solved by baryonic physics effects, most notably the effect of feedback
processes that selectively eject low angular momentum baryons and produce cores in low-mass dark halos via impulsive heating
of the dark matter cusps (\citealt{GerhardBinney}; \citealt{ReadGilmore}; \citealt{Governato04}; \citealt{governato10}; \citealt{PontzenGovernato}). However,
unsolved issues remain in the number counts of dwarf galaxies, both among satellite galaxies of large spirals, such in our Local Group, 
and in the field. Indeed, while the dearth of faint satellite galaxies, hosted by halos with $V_{vir} < 20$~km/s, can be explained by
the combined effect of reionization, stellar feedback and environmental processes (\citealt{Bullock}; \citealt{brooks}; \citealt{diCintio}), 
it has been pointed out that there is still a possible excess of massive satellite halos with $V_{vir} > 20$~km/s relative to 
the number of observed bright dwarf galaxies \citep{TBTF}, which also appear to have lower central densities than predicted
by CDM. Squelching by reionization cannot provide a simple solution in the latter case since gas will be retained 
within hosts of this mass \citep{grumpy}, which actually ought to be even more massive before infall \citep{Mayer10}.
Cosmological hydrodynamical simulations, however, suggest that feedback before infall may modify the DM density profiles enough to
reduce the central densities of satellites (\citealt{brooks}; \citealt{Wheeler15}), while recent simulations embedding high-resolution
models of satellites within cosmological MW-sized halos do show that tidal stripping and stirring of satellites with such previously modified
central DM profiles may have a strong effect on the resulting satellite population, possibly eliminating the ``massive failures" (\citealt{Tomozeiu}; \citealt{Tomozeiub}). Yet, also {\it in the field} a dearth of dwarfs with $V_{vir} \sim 40-60$~km/s
has been noted relative to CDM prediction \citep{Klypin} which is harder to explain since one cannot rely on the combination of feedback
and environmental effects.


In order to seek alternative solutions to these small scale problems recently, there has been revived interest in other dark matter
models such as self-interacting dark matter (SIDM) and 
warm dark matter (WDM) model, or better in models that give rise to a truncated power spectrum of density fluctuations at scales 
close to those of dwarf galaxies (\citealt{sommer}; \citealt{lovell}; \citealt{Weinberg15}). 
\citet{Fry15} have used cosmological hydrodynamical simulations to show that SIDM does not modify the central density of the
dark halo in dwarfs with peak velocities less than $30$~km/s, a range where also baryonic feedback effects are inefficient, while above
that baryonic feedback dominates and leads to results almost identical to CDM, at least for a fixed 2 cm$^2$\,g$^{-1}$ SIDM cross section.
It remains to be seen how SIDM models with a velocity-dependent cross section would behave. 

By construction WDM models reduce
the abundance of dwarf galaxies, possibly up to the scale of the ``massive failures" as long as the particle rest mass energy is high enough
(above 1~keV).  Here we consider WDM models in the $1-3$~keV range. Compared to CDM, WDM particle is much lighter, therefore has more significant free-streaming. Popular candidates for WDM are the 
gravitino \citep{gorbunov} and the (right-handed) sterile neutrino \citep{Drewes}, with a Fermi-Dirac-like momentum distribution, which can yield the desired 
cut-off of the power spectrum at small-scales. These are models that follow in the category of {\it thermal relics}, and this is
the category that we will always implicitly assume in this paper.
Most stringent constraints on WDM mass are given by the Lyman-$\alpha$ forest with $m\geqslant3.3$~keV to a 2-$\sigma$ confidence level \citep{Viel}. The number counts of dwarf galaxies 
yield also $m\geqslant2.3$~keV \citep{Kennedy}, at 2-$\sigma$ confidence level too, but with further uncertainties coming from the mapping 
between dark matter halos and baryons.

\citep{lovell} performed WDM pure N-body zoom-in simulations of MW-sized halos assuming a sterile neutrino model in the range $1.4-2.3$~keV, finding that 
they can naturally avoid the ``too-big-to-fail" problem highlighted by \citep{TBTF}. However, in order to be a credible model for galaxy formation,
WDM needs to pass a number of tests. Among these are the rate and timing of assembly of the baryons inside galaxies, which are reflected in their
star-formation histories and final stellar masses. As in WDM galaxies tend to form later than in CDM the timing of the assembly of their baryonic components will indeed be 
affected, but to what extent and whether or not this is measurable with some clear diagnostic is not yet firm.
\cite{governato} have begun to address these important aspects using a small set of zoom-in
hydrodynamical simulations of field dwarfs. They
have shown that the later formation time can leave an imprint in the star-formation histories (SFHs) of dwarfs and even more in the properties of the
stellar populations traced with Color-Magnitude Diagrams. Inspired by these recent numerical results, we attempt a novel test of WDM models in the
context of galaxy formation. The test uses star-formation histories of dwarf galaxy satellites to infer a range of possible formation times, and
compares them with the formation time expected in different WDM scenarios as opposed to CDM. We use dwarf galaxies which are known to 
be highly dark-matter-dominated, hence appropriate nearby objects to probe more directly different dark matter models.
Moreover most dwarf spheroidals show no currently active star-formation sites and all of them have a substantial, often highly 
predominant, population of very old stars (e.g. \citealt{Gallart15}), allowing to probe directly the early assembly history of such objects,
which is the thrust of the method that we propose here.

We compare the formation time of the dwarf dark halo inferred in CDM and WDM models by our method and 
the time at which the galaxy has formed 90\%, of its stars, namely the bulk of its stellar component. This is a timely analysis owing to the
improved accuracy of SF histories of local dwarfs by means of HST-based high-quality color-magnitude diagrams (\citealt{weisz}; \citealt{weiszAndII} and the
papers of the LCID collaboration \citealt{Bernard08}; \citealt{Aparicio}). We determine if a dwarf SFH to be consistent or not with a certain WDM scenario by requiring
that its WDM halo cannot form later than the bulk of its stars. 

We note that \cite{Calura14} have also used stellar ages to constrain WDM. However they adopted
a different approach, namely they used the ability to reproduce the
luminosity function of low-mass galaxies as a way to discriminate
between different dark matter models. By using stellar ages of  Local Group dwarfs we adopt a much smaller
sample of objects but free ourselves to uncertainties in the completeness of samples, crucial for the
luminosity function at the faint end. Also focusing on local galaxies for which any property, including
stellar ages, is known with much better accuracy than anywhere else, yields in principle more
robust constraints.

The paper is organized as follows. In Section \ref{methods} we describe our methodology to determine the formation time (collapse redshift) of dwarf
galaxy satellites, highlighting the assumptions on which it is based. 
In passing we also discuss a self-consistency check on the infall mass assignment by means of the dynamical friction timescale. In Section
\ref{results} we present our main results concerning WDM models with particles having masses in the range 2-3~keV, focusing on the 
MW Ultra-Faint Dwarfs which we found to be the most constraining objects in our sample. In Section \ref{results1kev} we present our results to place constraints
on the 1~keV particles, this time using both MW Ultra-Faint Dwarfs and the classical dSphs. In Section \ref{caveats} we discuss the caveats of our
methods and in Section \ref{conclusion} we provide concise conclusions. An Appendix follows which shows the constraints on 2-3~keV models coming from
the SF histories of classical dSphs and a more thorough explanation of the dynamical friction argument that we use as a further check of the
range of plausible infall masses for a given dwarf galaxy.

\section{Methods}\label{methods}
\begin{figure*}[]
\includegraphics[width=0.48\textwidth]{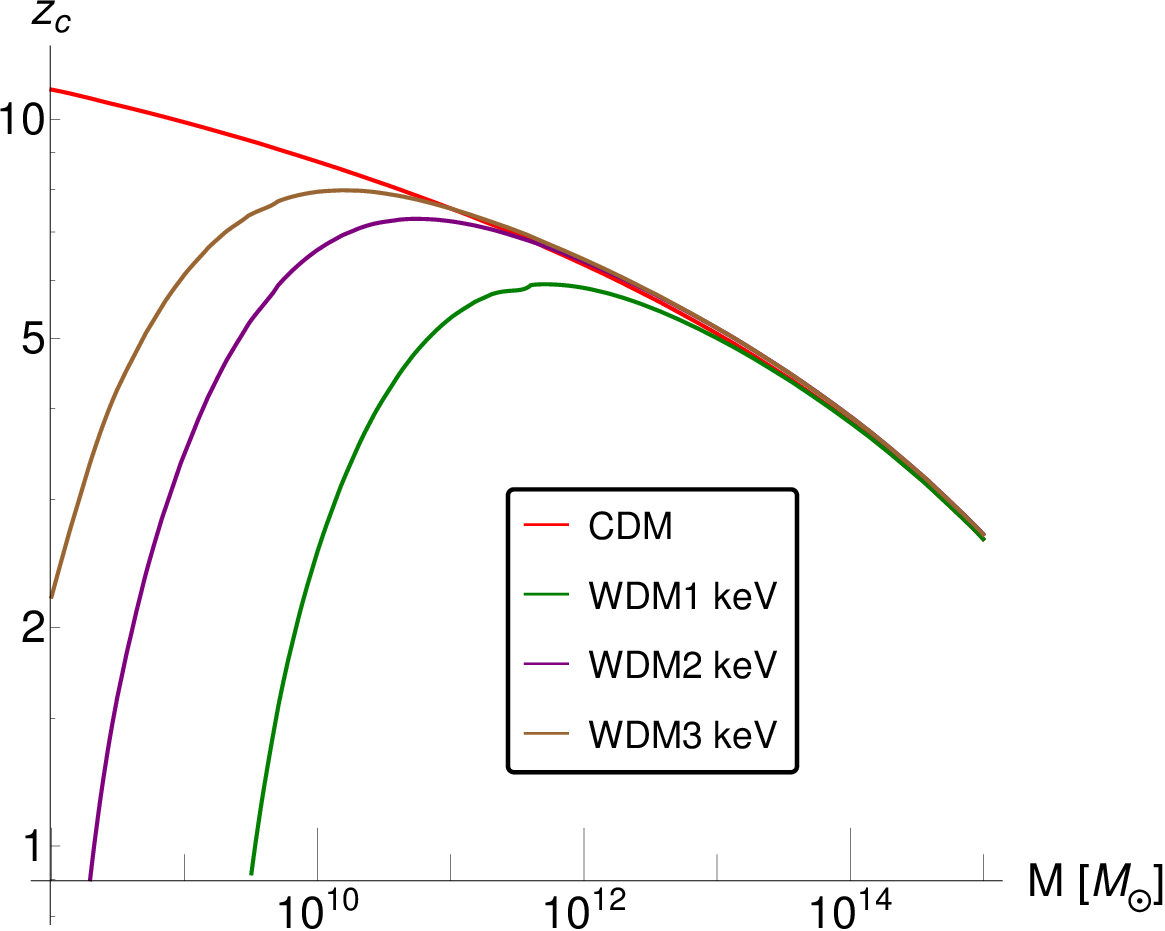}
\includegraphics[width=0.48\textwidth]{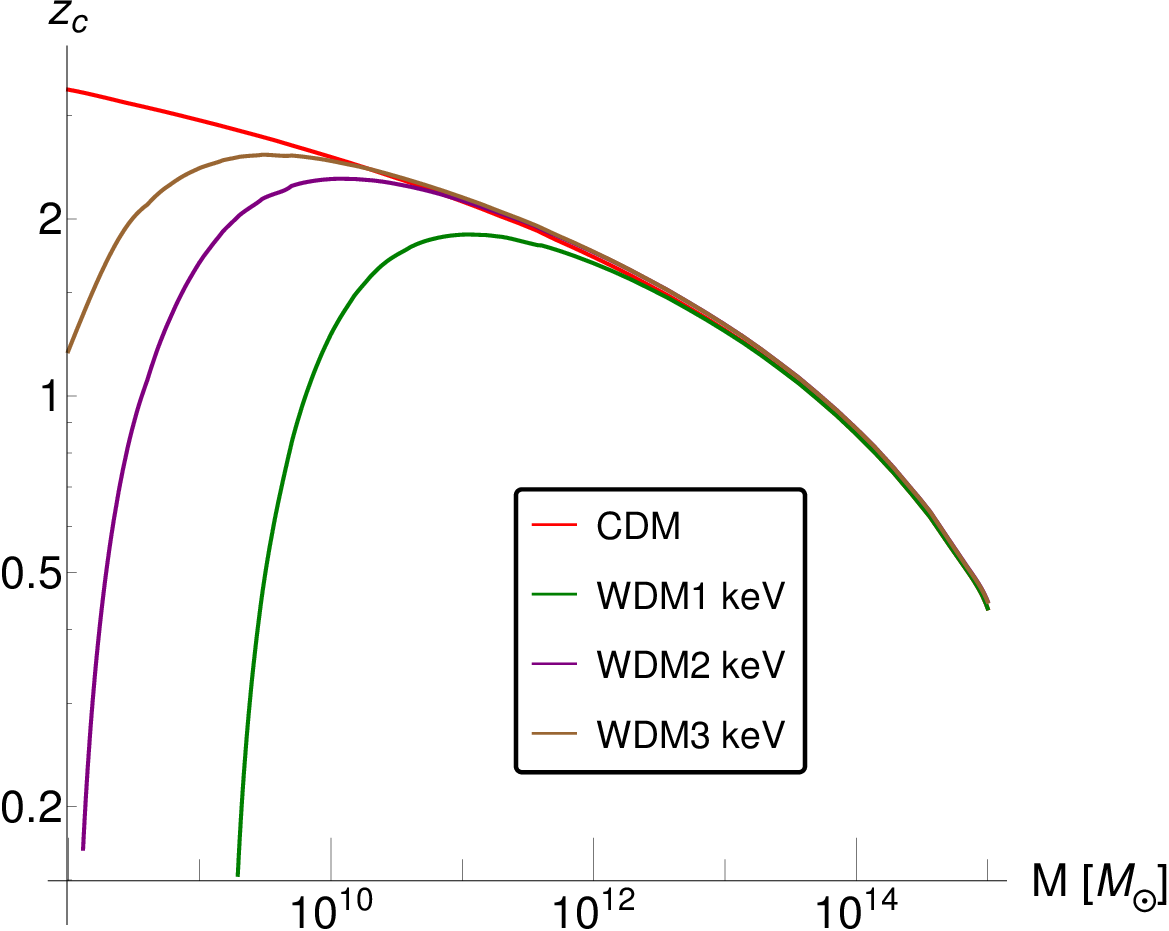}
\caption{Collapse redshift for a halo that accreted 4\% (left panel) and 50\% (right panel) of its mass $M$ at redshift 0 in different scenarii, where WDM$\chi$~keV means WDM with a particle mass $m=\chi$~keV. The curves are normalized to the Bolshoi simulations, which determine $z_c$ for the upper mass range $M=10^{11}-10^{15}~$M$_\odot$.\label{zcM}}
\end{figure*}
We aim to derive the collapse redshift of a dwarf satellite halo through the available kinematic data. 
In CDM we may exploit the $c-z_c$ relation to infer the collapse redshift; however this is excluded in WDM models since the relation is not monotonic anymore. We turn to an extended \citeauthor{Schechter} (EPS) model which yields $z_c-M$ curves. \cite{wolf} showed that the mass of the dwarf halos can be inferred via their kinematic data, whose method was developed further by \cite{BBK}. We start by briefly reviewing their approaches before turning on to EPS in WDM. We close this section with \cite{weisz} reconstruction of star-formation.

\subsection{Infall mass assignment}
\cite{wolf} infer the mass of the host halo of the Milky Way satellites from the 3D deprojected 
half-light radius $r_{1/2}$ and the mass enclosed within it $M_{1/2}\equiv M(r_{1/2})$, which can be solidly approximated by:
\begin{equation}
M_{1/2}\simeq \frac{3}{G}\sigma_*^2r_{1/2},
\end{equation}
where $\sigma_*$ is the stellar velocity dispersion, which is assumed to be flat near the half-light radius.
For most profiles, $r_{1/2}$ can be approximated by $r_h \simeq 3/4 r_{1/2}$, where $r_h$ is the (two dimensional) projected half-light radius. $r_h$ and $\sigma_*$ are taken from \cite{mc}. To a good extent, $M_{1/2}$ owes the largest contribution to the dark matter mass enclosed
within the half-light radius since dSphs are dark-matter dominated at all radii. Therefore we neglect the stellar mass and determine the total
mass of the dwarf by assuming that its total mass is that of an NFW halo whose enclosed mass $M(r)$ for $r=r_{1/2}$ matches
the observationally determined value of $M_{1/2}$. In doing that we have the freedom of choosing the concentration $c$ of the NFW profile.
Therefore, in practice we first assume a possible $M_{200}$ and then determine if there is a $M_{200}-c$ pair that fits the constraints by using \cite{maccio} fitting formula.
We note that it has been already shown that the NFW model is a good model to describe the dark matter distribution also in the case of a WDM
universe, see \cite{lovell}, therefore the same procedure can be repeated in both cases. The concentration values implied by the fitting procedures
are different in the two cases though, as WDM concentrations are found to be lower both by analytical models and simulations (e.g. \cite{Colin00}; \cite{bode}; dropping
below 10 \citep{aurel}.
\citeauthor{wolf} consider $M_{200}$
as a good proxy for the mass of the subhalo hosting the dwarf before infall into the primary halo
of the Milky Way or M31, which we will refer to as $M_{infall}$. Table \ref{cCDM} and Table \ref{cWDM} report the $M_{infall}, c$ pair of values adopted for the
CDM subhalo models and various WDM subhalo models considered throughout the paper and discussed later on.

\begin{table}[h!]
\begin{center}
\begin{tabular}{l|l|l|l|l}
$M_{200}~[M_{\odot}]$ & $3\cdot10^7$ & $3\cdot10^8$ & $3\cdot10^9$ & $3\cdot10^{10}$ \\ \hline
$c_{200}$ & 18.2 & 14.5 & 11.6 & 9.2
\end{tabular}
\caption{Mean value of CDM halo concentrations for a given $M_{infall}$, assumed to be $M_{200}$, in Wolf et al. procedure. Values are taken from \cite{maccio} fitting formula. \label{cCDM}}
\end{center}
\end{table}
\begin{table}[h!]
\centering
\begin{tabular}{l|l|l|l}
$M_{infall}~[M_{\odot}]$ & $10^8$ & $10^9$ & $10^{10}$ \\ \hline
$c$ in WDM2 & 8 & 10 & 12 \\ 
$c$ in WDM3 & 12 & 14 & 14.5 \\
\end{tabular}
\caption{Typical halo concentrations in warm model numerical simulations \citep{aurel}, where WDM$\chi$ indicates a warm particle of $m=\chi$~keV.\label{cWDM}}
\end{table}

Note that the dark matter profile of dwarf galaxies is likely flattened away from the NFW profile due to baryonic feedback effects (\citealt{governato10}; \citealt{Teyssier13}). We will discuss
later, in section 5, how this could affect our conclusions in light of recent work based on hydrodynamical simulations. 
We are also aware that dwarf galaxies lose mass due to tidal mass loss
during the interaction with the primary halo. This does not reduce simply to tidal truncation of the halo inward to the nominal virial
radius $R_{200}$, which can be described with a simple exponential cut-off \citep{Kazantzidis04} and would not affect the
determination of $M_{infall}$ but includes also 
the effect of repeated tidal shocks which can strip the subhalo much further inside, depending on initial orbit 
and halo concentration (\citealt{Taffoni}; \citealt{Zolotov12}). The latter effect can reduce the peak circular velocity, which essentially
corresponds to a reduction of the enclosed mass at radii of order the scale radius of the NFW halo, which in turn is of order
the size of the luminous component of dSphs \citep{Kazantzidis11}. This reduction can be mild or quite strong depending
on the orbit of the subhalo, but also depending on whether or not the inner profile of the subhalo develops a core-like distribution
due to baryonic effects, an aspect that has emerged from simulations only recently \citep{Kazantzidis13}. Recent cosmological
simulations that can model the combined evolution of the stellar and the dark matter components of dwarf galaxy satellites of Milky Way-sized
halos, but only for a limited set of objects, show that in some cases $M_{infall}$ can be underestimated by up to a factor of 10 with 
such procedure. The largest offsets occur when the inner halo distribution is shallower than inferred from the NFW profile \citep{Tomozeiu}.

To overcome at least partially these effects we exploit the results of \cite{BBK}, which are based on cosmological dark matter only simulations.
These simulations of course still miss the direct mapping between radii and masses of the luminous component of the dwarf and those of its
halo, and neglect baryonic effects that can change the inner dark matter density slope. We will provide a discussion of the remaining
caveats at the end of the paper. 
Using numerical simulation from the Aquarius project, they compute properties, such as the infall mass, of DM suhhalos that are consistent with the dynamics of the brightest dSphs. They assume that the simulated subhalos are a representative sample for $\Lambda$CDM simulations. Though they worked 
only with $\Lambda$CDM simulations to estimate the infall mass, studies of subhalo properties carried out in WDM models \citep{ander}
suggest that the only important difference in applying this matching procedure would be in the halo concentration parameter 
since for a \textit{given} CDM halo, WDM produces a slightly smaller halo and subhalo concentration than in CDM. 

Furthermore, the analysis of \cite{BBK} is performed only for classical dSphs, which automatically selects fairly high $M_{infall}$, often above
the filtering mass of WDM models (and always well above the free-streaming mass). This means that such analysis is not expected
to be constraining for WDM models since statistically there will be enough subhalos to find $M_{200}-c$ pairs that fit the observational
constraints. Nevertheless we carry out the matching and subsequent analysis of the star-formation history constraints for the WDM models using the infall masses determined by \cite{BBK} and report this in the Appendix \ref{MW}. 
For the two other sets of dwarfs, which are the Andromeda ones and the ultra-faint dwarfs of the Milky Way, which are potentially much more
constraining because they include objects with much lower present-day estimated stellar and dark halo mass, the only available method is 
the \citeauthor{wolf}'s approach, which yields only an approximate estimate of the possible $M_{infall}$. The main analysis
presented in this paper is based on the latter method.

\subsection{Collapse redshift via an extended Press-Schechter formalism}
The Bullock model is unable to reproduce the turnover in the concentration-mass relation. \cite{aurel} showed that it can be improved with an extended Press-Schechter formalism and by requiring the average collapse redshift of halos that survive until today. $z_c$ is defined as the time when the halo has accreted a fraction $F$ of its final mass $M$.

The average growth factor of all the progenitors can be straightforwardly derived:
\begin{equation}
D(z_c)=\left(1+\sqrt{\frac{\pi}{2}}\frac{1}{\delta_{c,0}}\sqrt{S_\chi(FM)-S_\chi(M)}\right)^{-1},
\end{equation}
where $S_\chi$ is the variance for a given $\chi$ DM scenario. $D(z)$ can be inverted to find the collapse redshift. \cite{aurel} showed that the slope of the curves $z_c-M$ fits results from CDM simulations up to a normalization. We use here the Bolshoi simulation results as quoted by \cite{bosch}, where two criteria are presented: the collapse redshift by when the halo accreted into 4\% and half of its final mass. The curves are illustrated in Figure \ref{zcM}.

To derive the power spectrum and hence the variance $S_\chi$, we use the linear transfer functions computed with the \texttt{CLASS} code for a 2 and 3~keV WDM particle with a Fermi-Dirac-like angular momentum distribution and the fitting function of \cite{Viel} for the 1~keV model. For the cosmological parameters, we use the values obtained by the Planck collaboration, i.e. $H_0=68.14$, $\Omega_m=0.304$, $n_s=0.9$, $\sigma_8=0.827$.

Once the total mass $M_{infall}$ of each galaxy dwarf is known, we derive the collapse redshift with the curves drawn in Figure \ref{zcM} for the considered WDM model. 

\subsection{Dynamical friction constraints on infall mass\label{dfric}}

We also develop a separate argument that serves as a self-consistency check on the infall masses derived with the method just described.
This is important since, as we discussed, the mass inference in WDM 
is uncertain by nature. We determine the highest subhalo infall masses that we could assign for WDM models from the \citeauthor{wolf}'s approach 
and still satisfy the natural constraint that the dynamical friction timescale at infall has to be (sufficiently) 
longer than the time elapsed between infall and the present time. This condition is quantitatively expressed by requiring
that individual dwarf spheroidal or ultra-faint satellites have to end up at present-day galactocentric distances comparable
to those at which they are found today relative to the MW or M31.
Note the concentrations assigned in the standard procedures are on the high side
of those expected in WDM models, so we can instead start by assigning a typical concentration measured in WDM
simulations \citep{aurel}, which then yields halos that are 5 to 10 times heavier than in our default method.
This choice also matches well the extrapolation of the halo mass-stellar mass relation, see \cite{grumpy}, \cite{Behroozi}. By using a dynamical 
friction time estimate, we then computed the earliest infall time the galaxy could have, based on their current distance to the MW center.It turns
out that from this analysis alone, it is
not possible to exclude higher infall masses that would accommodate even a 1~keV model: only for an extreme parameter choice 
does the dynamical friction time scale become too short once one takes into account also the ralenting effect of tidal mass loss on orbital decay. 
On the other hand, the infall masses that we obtain based on our default method are within the range of those admitted by the
dynamical friction argument so in this sense the self-consistency check is successful.
The analysis is presented in Appendix \ref{DF}. 

\subsection{star-formation history}
It is only recently that extended star-formation histories have been determined, thanks to the high-quality color-magnitude diagrams (CMDs) produced by the Hubble Space Telescope (HST). Its instrumental uniformity has allowed \cite{weisz} to present the first consistent analysis for more than 40 dwarf galaxies of the Local Group. The star-formation histories are measured from CMDs using the maximum likelihood CMD fitting routine, \texttt{MATCH}. Not all the CMDs reach the oldest main sequence turn-off (oMSTO) but \citeauthor{weisz} argue that only the first epoch(s) of star-formation (SF) are unconstrained by their method. This property does not hamper our analysis since we are interested in the latest epochs of SF. 

For the ultra-faint dwarfs whose a CMD analysis has been made, i.e. Ursa Major, Coma Berenice, Hercules, Leo IV, Bootes I and Canes Venatici II, we use \citet{brown14} conclusion: the dwarf galaxies formed 80\% and 100\% of their stars by $z\sim6$ (12.8 Gyr ago) and $z\sim3$ (11.6 Gyr ago), respectively. They have used ACS/HST CMDs reaching well below the oMSTO. We compare their result with our reference \cite{weisz} when results are available in the Appendix \ref{SFH}. The values are in both methods in favour of an early SF stopping except for CVnII. We use \citeauthor{brown14} results because they are available for a more extensive sample of the ultra-faint dwarfs.

For each dwarf galaxy whose respective data are available, we compare the collapse redshift with the redshift at which the galaxy formed 90\% of its stars, except for the ultra-faint dwarfs where we use the time by when they formed 80\% to 100\% of their stars as this is more consistent with the available stellar age resolution in the observations.

\section{Constraining 2-3~keV WDM models: the SF histories of MW ultra-faint dwarfs}\label{results}
In this Section we report the main results of our work, which concern the
interesting constraints on WDM scenarios obtained by using the star-formation histories and 
expected collapse times for the ultra-faint dwarfs (UFDs) of the Milky Way. 
In the Appendix \ref{MW} we show also analogous results for 8 bright MW dwarfs and a set of Andromeda satellites 
for which SF histories are available which, as we anticipated, are not providing strong constraints on WDM scenarios.

\begin{figure}[h!]
\includegraphics[width=0.48\textwidth]{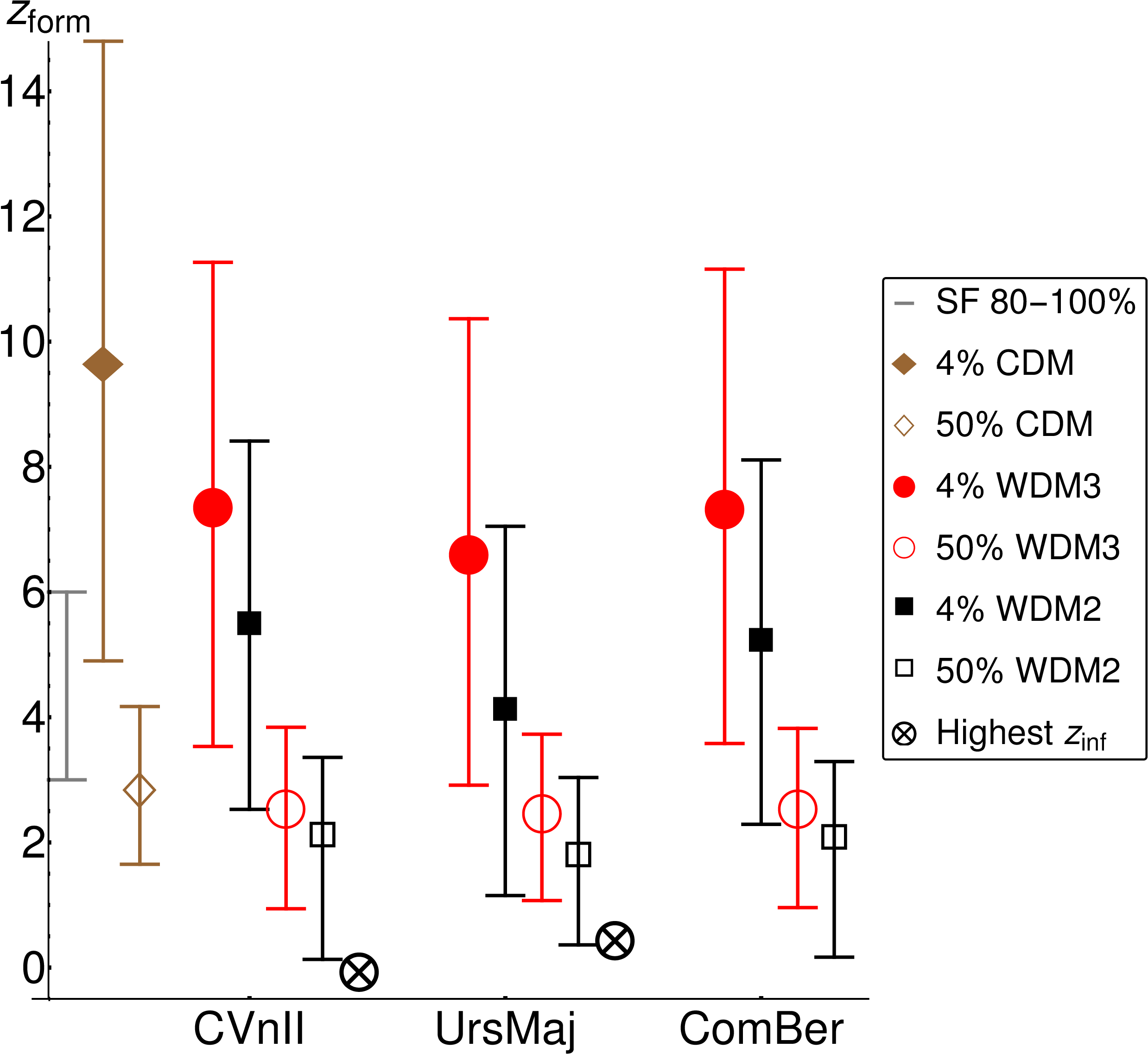}\caption{Collapse redshift distribution for the three massive ultra-faint dwarfs of the MW indicated on the $x$-axis, for the different dark matter models and collapse redshift criteria indicated in the upper right corner. The redshift by when the galaxies formed 80\% and all of their stars is despicted left in green and valid for all dwarfs of the sample. We show both collapse redshifts, by when 4\% and 50\% of the final mass were formed, in the different warm models, indicated by WDM$\chi$ where $m=\chi~$keV. We also show the highest infall redshift coming from the dynamical friction criterion discussed in Section \ref{dfric} for Canes Venatici II and Ursa Major I; Coma Berenices, being closer to the galactic center, should have fallen at $z>10$, which would be out of the frame and hence less relevant for our analysis. The error bars show the errors on the infall mass inference and the 2-$\sigma$ scatter of $z_c$ at a given mass. The shift in the $x$-axis has been arbitrary chosen for the purpose of illustration.}\label{summaryMWUFD}
\end{figure}

\begin{figure}[h!]
\includegraphics[width=0.48\textwidth]{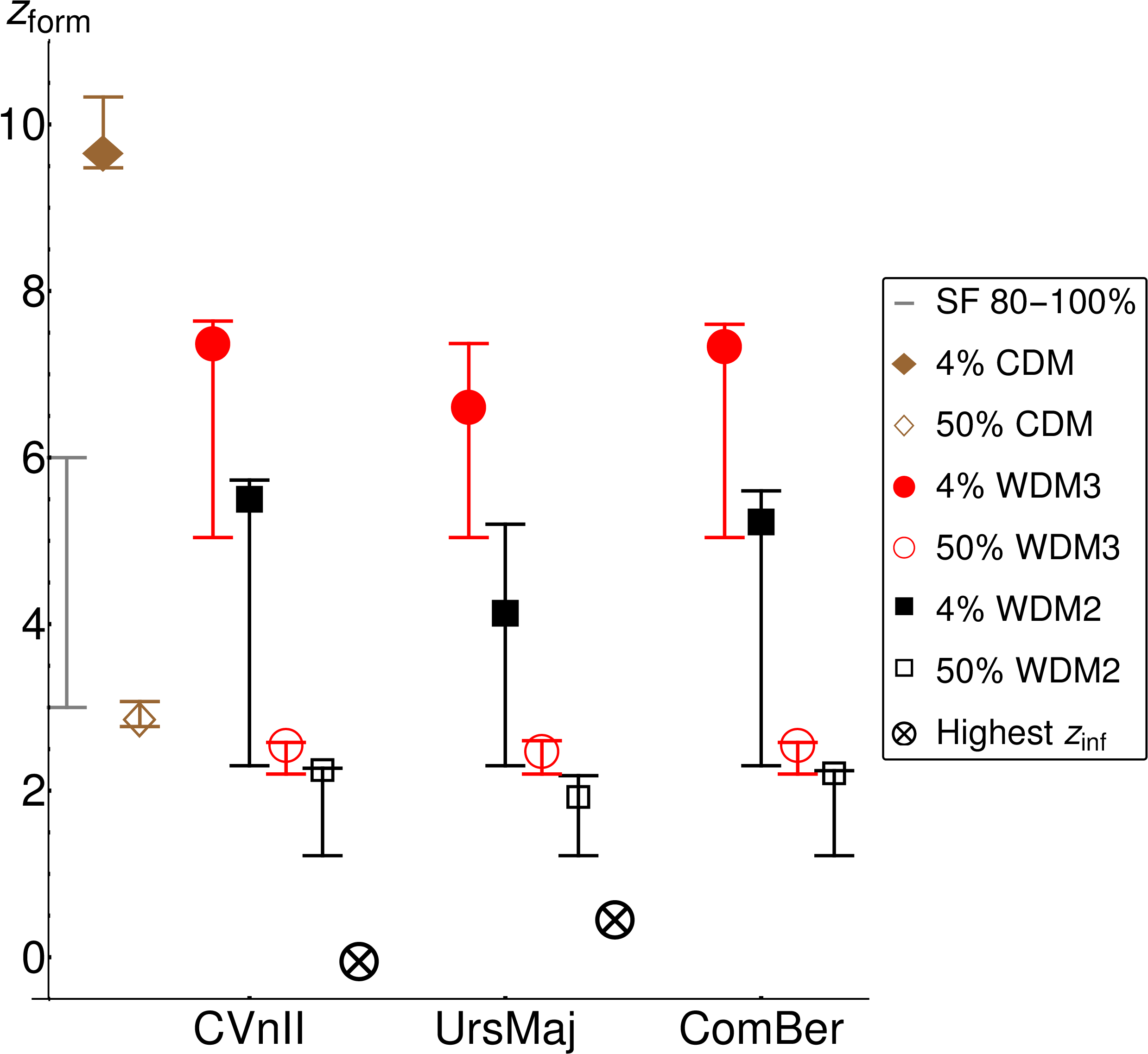}\caption{Collapse redshift distribution for the three massive ultra-faint dwarfs of the MW indicated on the $x$-axis, for the different dark matter models indicated in the upper right corner. The results are reported from Figure \ref{summaryMWUFD}. The error bars only show the errors on the infall mass inference, i.e. without the 2-$\sigma$ scatter of $z_c$ at a given mass.}\label{summarysansscatterMWUFD}
\end{figure}

\begin{figure}[h!]
\includegraphics[width=0.48\textwidth]{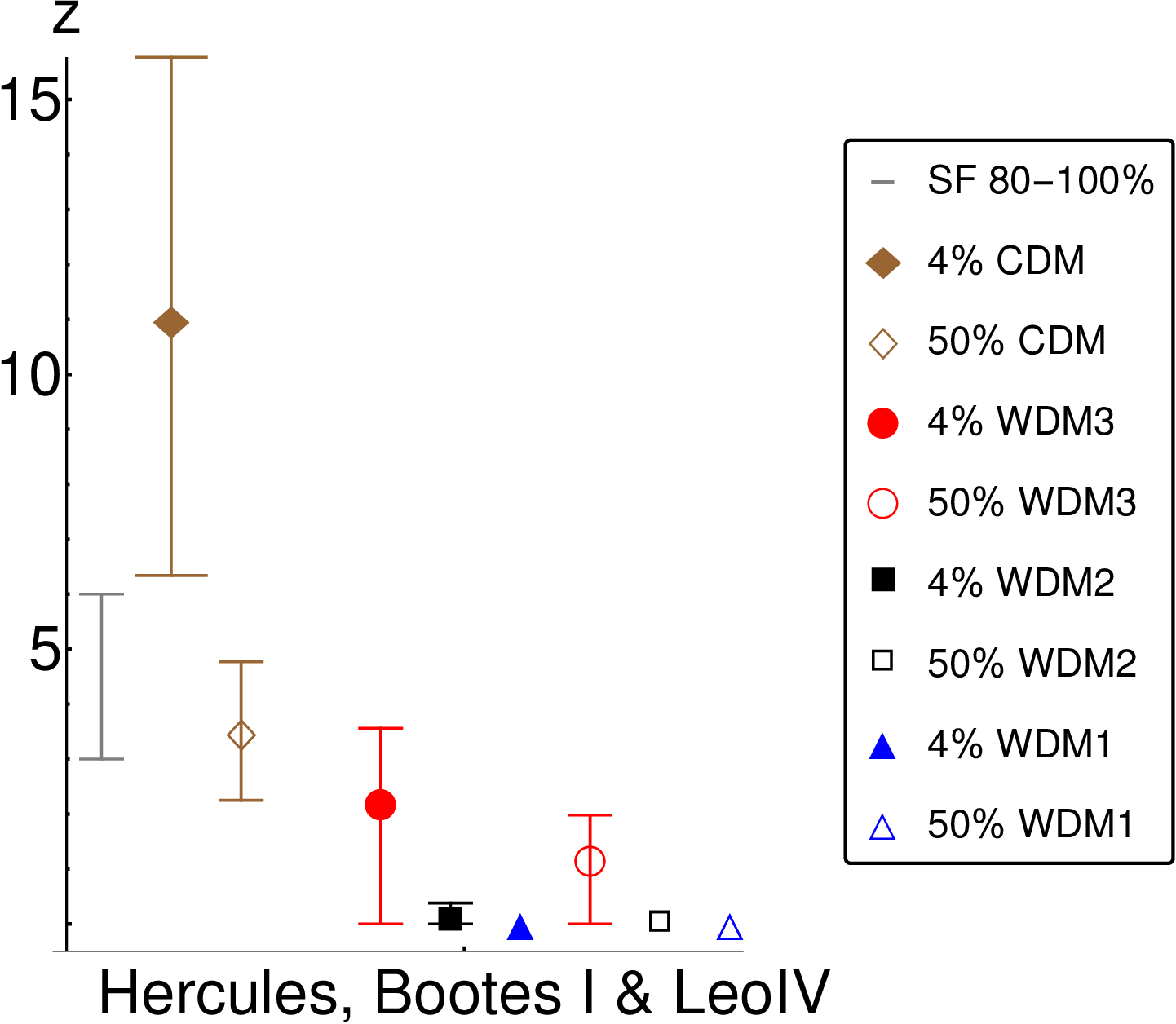}\caption{Collapse redshift distribution for the three light ultra-faint dwarfs of our sample, for the different dark matter models indicated in the upper right corner. We assume here a generic infall mass of $10^8~$M$_\odot$ for the three dwarfs, following our argument of tidal loss. The redshift by when the galaxies formed 80\% and all of their stars is despicted left in green. The errors correspond to the 2$\sigma$ confidence interval, which corresponds to the errors on the infall mass inference and the 2-$\sigma$ scatter of $z_c$ at a given mass. The constraints coming from the dynamical friction are weaker for light dwarfs, $z>10$, hence less relevant and not shown here. The shift in the $x$-axis has been arbitrary chosen for the purpose of illustration.}\label{MWUFD8}
\end{figure}

Our analysis is restricted to the ultra-faint dwarfs which have SF histories based on CMD diagrams; Ursa Major, Coma Berenice, Hercules, 
Leo IV, Bootes I and Canes Venatici II. Results are shown in Figure \ref{summaryMWUFD}, Figure \ref{summarysansscatterMWUFD} and Figure \ref{MWUFD8} where, for each dwarf, we show 
the collapse redshift inferred from the infall mass assignment procedure of Wolf et al. and the expected time for the bulk of the star-formation to have taken place, both with the (significant) error bars. 
The star-formation histories are taken from \citep{brown14} and are valid for all dwarfs; we show it in green at the left side of the figure. Their CMDs are nearly indistinguishable, which implies that their star-formation histories are largely synchronized, with the agreement estimated at the level of 1 Gyr.

For all the dwarfs we show the collapse redshift computed for the 2 and 3~keV WDM scenarios and for the CDM scenario. The 1~keV WDM scenario,
which is already nearly ruled out by other types of constraints (e.g. the Lyman alpha forest) will be discussed separately in the next section.
Note that we use two definitions of collapse redshifts, namely the redshift at which the subhalo has acquired, respectively, 4\% and 50\% of its final mass, both derived with the methodology described 
in Section \ref{methods}, with the infall mass determined via \citeauthor{wolf}'s approach. The two cases are shown in Figure \ref{zcM} for the various
scenarios.
While the 50\% criterion might seem the most sensible
and representative, we note that collapse redshift refers to the halo while the star-formation history refers to the baryons. Since it is conceivable
that the baryonic component of the galaxy assembles in the inner region of the subhalo, and since even present-day {\it field} dwarf galaxies
appear to reside within a few percent of the virial radius of the halo (e.g. \citealt{littlethings}), it is equally reasonable
to argue on favour of the $4\%$ criterion as more representative as most of the halo mass can be gathered later than the assembly of the
baryonic galaxy.

The error bars correspond to the raw $1\sigma$ errors on the inferred $M_{infall}$ (taken from \citealt{BBK} for the classical MW satellites, 
based on $M_{1/2}$ for all other satellites) plus the $2\sigma$ scatter of the collapse redshift at a given infall mass. The spread in concentration, which is what
determines the error in the collapse redshift is that determined from CDM simulations, $\Delta \log(c)=0.14$, as quoted by \cite{wolf}. 
We assume such scatter to be relevant also for the WDM scenarios, although this would require a systematic numerical study for verification.

For clarity, Figure \ref{summarysansscatterMWUFD} only shows the errors due to the uncertainties in the infall mass inference. In this plot the errors bars are larger for the 4\% formation redshift, because the dependence of collapse redshift on mass turns out to be much steeper using the first definition relative to the second one (see Figure \ref{zcM}). 

Among our 6 UFDs, our infall mass assignment procedure yields three that are hosted in very light subhalos at infall $M_{infall}\sim10^7$. 
These are Hercules, Leo IV and Bootes I. At that value this would imply that the infall mass is already close to the free-streaming mass of the 2~keV model, 
$4\cdot10^6$~M$_\odot$, hence would be strongly suppressed in such a case. The free-streaming mass of the 3~keV model is slightly lower $\sim10^6$, but
still this model would be marginally consistent. However, excluding the two models already at this stage would be likely erroneous since these dwarfs,
if they are hosted in the smallest halos, are also the most likely to be strongly affected by tides. Using cosmological simulations
augmented with high-resolution dwarf galaxy models, \cite{Tomozeiu} have indeed
shown that an object like Bootes I could originate from tidal stirring of a much more massive disky dwarf falling into the Milky Way halo
at $z > 2$, especially if the halo profile was made core-like in the inner region due to baryonic effects.
In those simulations the progenitor of Bootes I could have had an infall mass $M_{infall} > 10^8 M_{\odot}$. Therefore,
we decide to carry out the analysis using $M_{infall} > 10^8 M_{\odot}$.
The results are shown in Figure \ref{MWUFD8} and are perhaps the strongest in this paper.
While the interpretation of the results, as expected, are clearly different when using the 4\% or the 50\% criterion, the bottom line is that these
three dwarfs alone appear to rule out the 2~keV model and render rather marginal also the 3~keV model (excluded at $2 \sigma$ for the 50\% criterion, not ruled out with the 4 \% criterion).

For the other three UFDs the resulting constraints are less stringent and strongly depend on whether one considers the 4\% or the 50\% of the infall mass
as collapse redshift criterion. The results are shown in Figure 2 and Figure 3. For the 50\% criterion the formation time inferred in the 2, 
and even in the 3~keV models is very difficult to reconcile with their SF history even when accounting for errors. We conclude that the ultra-faint dwarfs 
almost exclude the 2 and 3~keV models if we use the 50\% criterion. Instead if we use the 4\% criterion as an upper limit, we observe that the formation time
inferred from the SF history and the collapse redshifts can overlap within the errors. We conclude that, while the 4\% criterion might not exclude 
the 2~keV model, it still favours $m\geqslant2~$keV.

In order to highlight the quantitative difference between possible collapse redshifts for these three dwarfs in CDM and WDM scenarios, we
show the results for these three dwarfs for the two scenarios in Figure \ref{summarysansscatterMWUFD}, without the 
scatter due to the infall redshift uncertainty caused by the error on the concentration (this is by construction the same in WDM and CDM
scenarios).
 As a comparison, the mean collapse time in the CDM scenario is shown on the left in brown. 
This value is always in agreement with the star-formation criterion, but arguably at the borderline. However, we expect this to depend on the
value of the concentration adopted as a reference. Since in CDM there is no intrinsic limit on how early a subhalo could collapse in principle one is allowed
to postulate that UFDs are simply a population of objects biased to form very early, and hence with concentrations systematically higher than average.

Therefore we argue that in the CDM scenario there always exists a combination of parameters that fit the star-formation history of the galaxy.
 Let us assume the dwarf halo to be much lighter at infall, for example $10^7$~$M_\odot$. This would still match the constraints
from the Wolf et al procedure and would have $c \sim 25$ based on simulations
\citep{aurel}, which would then yield a collapse redshift $z= 10$ for the 4\% criterion \citep{zhao}. 
This is clearly early enough to accommodate any of the SF histories of UFDs.

The constraint on infall redshift coming from the dynamical friction timescale is shown in Figure 2 and Figure 3 for the case in which the UFDs were hosted in heavy subhalos before infall, namely $M_{infall} = 10^{10}~M_{\odot}$. Indeed at such high infall masses the collapse redshift would be at the upper end of the error bars shown in the same Figures, making the WDM
models marginally consistent with the timing of star-formation for the 4\% criterion. The resulting infall redshift in this case is $z<1$. This reveals a potential tension with
the fact that the same UFDs appear to have stopped forming stars at $z > 2$. Due to such low
infall redshift it is impossible to explain the truncation of star-formation via environmental mechanisms such
as ram pressure stripping and tidal mass loss 
\citep{Mayer}. One cannot invoke reionization either to stop star-formation at $z > 2$ because that would be 
only for $M_{vir} < 
10^8~M_{\odot}$ (e.g. \citealt{Susa}; \citealt{Kaurov}), namely below the masses considered
here. Hence the star-formation history of such galaxies would be puzzling in a WDM scenario if they
have relatively large masses, unless some 
non-conventional explanation is required,
such as a possible assembly bias effect aided by the local ionizing background of the primary galaxy 
(see e.g. \citealt{Gallart15}). This suggests they are difficult to fit with WDM without violating some
constraints.

\section{Constraints on the 1~\textrm{keV} WDM model from Sf histories of LG dSphs}\label{results1kev}
\begin{figure}[h!]
\includegraphics[width=0.48\textwidth]{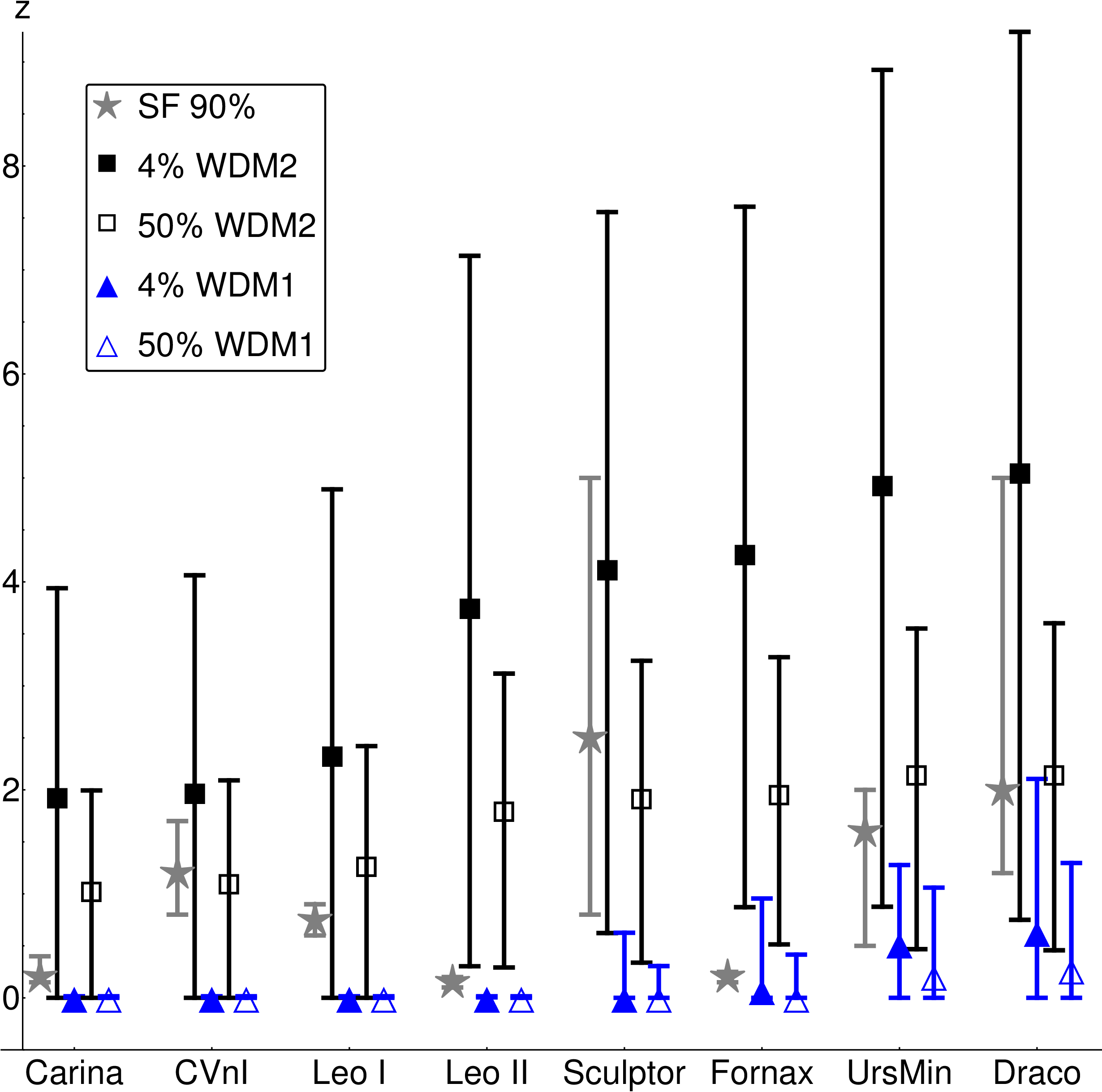}\caption{Collapse redshift distribution for the bright classical dwarfs of the MW, for the different dark matter models indicated in the upper left corner. For each dwarf, we also show the redshift by when the galaxy formed 90\% of its stars left in green. The error bars show the 2$\sigma$ confidence interval, which corresponds to the errors on the infall mass inference and the 2-$\sigma$ scatter of $z_c$ at a given mass. The shift in the $x$-axis has been arbitrary chosen for the purpose of illustration. }\label{MW1keV}
\end{figure}

\begin{figure}[h!]
\includegraphics[width=0.48\textwidth]{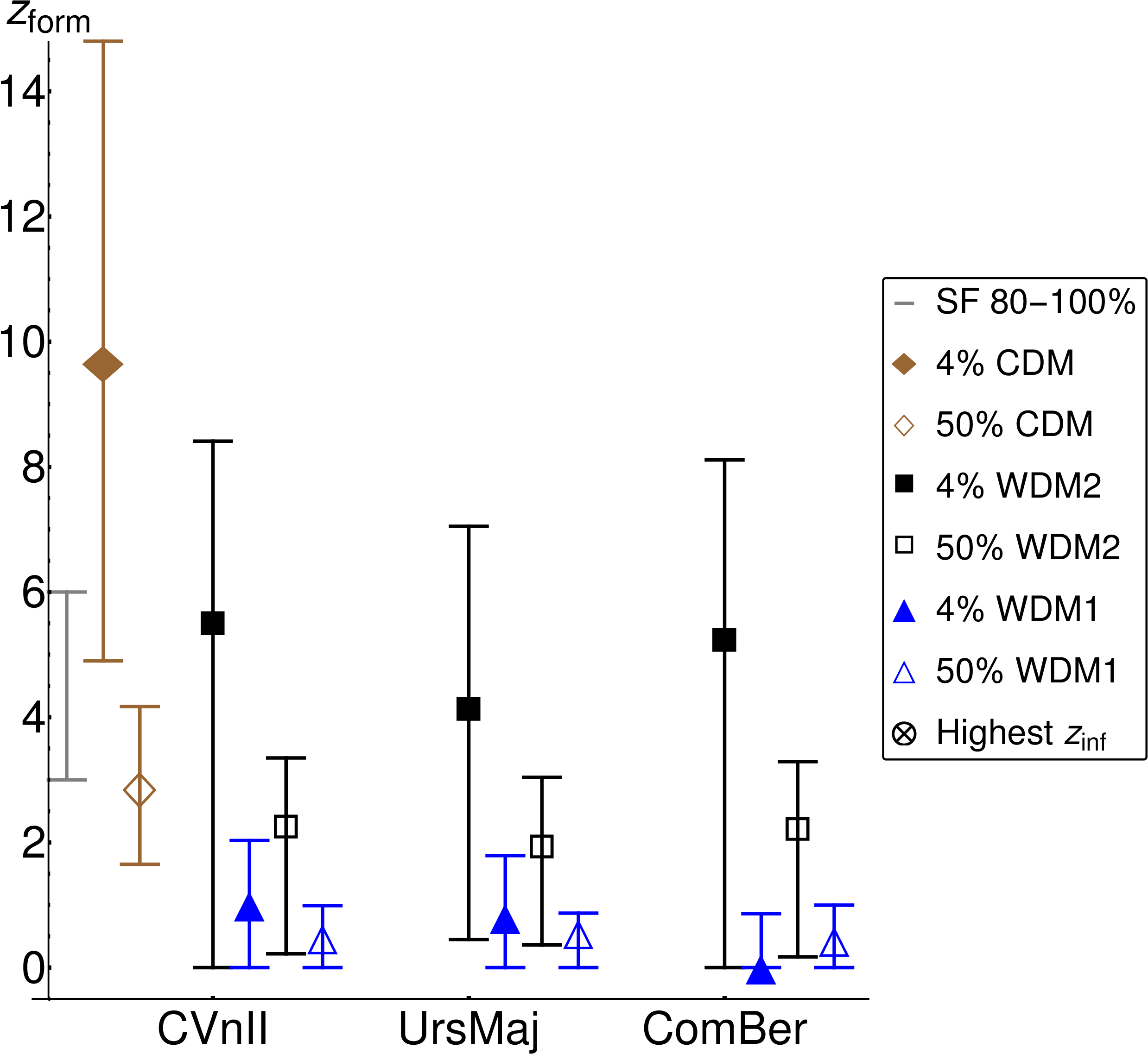}\caption{Collapse redshift distributiont for the three massive UFDs of the MW, for the different dark matter models indicated in the upper right corner. The redshift by when the galaxies formed 80\% and all of their stars is despicted left in green and valid for all dwarfs of the sample. The error bars show the 2-$\sigma$ confidence interval, which corresponds to the errors on the infall mass inference and the 2-$\sigma$ scatter of $z_c$ at a given mass. The shift in the $x$-axis has been arbitrary chosen for the purpose of illustration.}\label{MWUFD1keV}
\end{figure}

In order to estimate how strong is the constraining power of our method compared to other more conventional ones,
we repeat the analysis of the classical and ultra-faint dwarfs of the Milky Way with the 1~keV model. This is a WDM model that is 
indeed already nearly excluded by the Lyman-$\alpha$ forest (e.g. \citealt{Viel}). 
Figures \ref{MW1keV} and \ref{MWUFD1keV} compare the time needed for 90\%, and 80-100\% respectively, of the stars to be formed, with the collapse redshift in a WDM cosmology with a mass of 1~keV. The figures also show the expected collapse redshift in a 2~keV model, allowing for the comparison of the achieved accuracy of the two cosmogonies. Here the error bars show the uncertainties on the mass inference plus the 2-$\sigma$ scatter of $z_c$ at fixed mass.

For all dwarfs of the sample, the 1~keV model is clearly excluded at 2-$\sigma$, if based on the 50\% criterion. For the 4\% criterion, Fornax, Ursa Minor and Draco do not contradict the star-formation time, but the errors are too large to exclude the 1~keV model at 2-$\sigma$. However, on our whole sample of 11 dwarf galaxies, only the 3 dSphs Fornax, Ursa Minor and Draco exhibit significant uncertainties. We deduce that the 1~keV model is excluded by our approach. This joins other, independent, constraints on WDM, such as those given by the Lyman-$\alpha$ forest and the count of satellites.

\section{Caveats}\label{caveats}
One of the main weaknesses of our approach is its high sensitivity to the infall mass, which is due to the abrupt turnover in the formation 
redshift-mass relation in the WDM scenario (Figure 1). Our method is thus as accurate as the determination of the infall mass. Some of the
errors associated with the latter have been taken in to account in our analysis, such as the variation due to variation in the assumed
subhalo concentration. We have also considered to some extent the fact that the concentration itself might have been reduced by tidal shocks, making values
at the lowest end of those expected based on the mass-concentration relation found in numerical simulations absolutely plausible, a notion
that we have implicitly exploited when we assumed a higher infall mass for the 3 faintest UFDs. But there are more caveats.

Baryonic effects, completely neglected in our mapping of the infall mass from present-day dwarf properties, may affect the way we associate an infall mass to each galaxy when we follow the Wolf et al. procedure.

\cite{brook14} quantified this by adding an individual ratio $M_*/M_{halo}$ to the NFW profile and asking how that would modify the mass distribution
taking into account the effect of feedback associated with a given $M_*$ assembled in the halo, calibrated with a set of numerical hydrodynamical simulations of dwarf galaxy formation.

Their corrected profile generally associates galaxies to higher infall masses than the standard NFW profile owing to the same effect that we
have considered for the tidal shock argument -- the central density is reduced which corresponds effectively to lowering the concentration
of the halo and therefore increasing the value of $M_{infall}$ that can be assigned. We estimate that on average the assigned infall
mass would increase by a factor of 10, although large fluctuations can be expected on a case-by-case basis as there is not a simple
linear relation between the flattening of the profile and other properties of the dwarfs such as the present-day stellar-to-halo mass.
In any case the effect goes in the direction of allowing an earlier collapse redshift in the WDM models.
As an example, we analyze the borderline cases of some MW classical dSphs: Sculptor would be hosted by a subhalo having an infall mass 
of $1.5\cdot10^{10}$ and Ursa Minor 
by a subhalo of $10^{10}$. Even the formation redshift for the 50\% criterion is now not problematic, becoming 2.39 in the 2~keV model, 
which is early enough, compared to the redshift when the galaxy formed 90\% of its stars, which is just below 2. 

\begin{figure}[h!]
\includegraphics[width=0.48\textwidth]{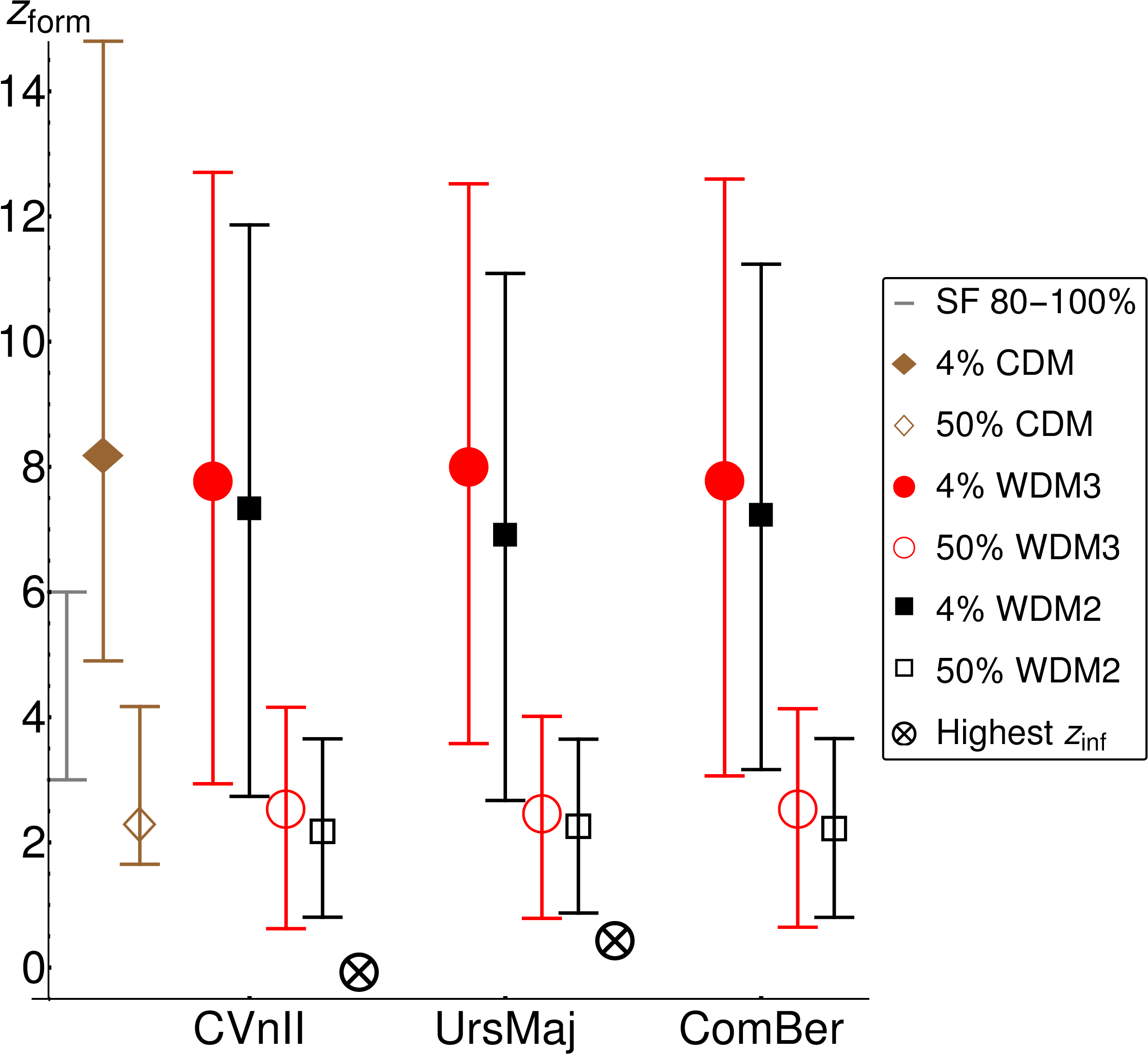}\caption{Collapse redshift distributiont for the three massive UFDs of the MW if their inferred mass would be 10 times heavier ($10^{10}~M_\odot$, for the different dark matter models indicated in the upper right corner. The redshift by when the galaxies formed 80\% and all of their stars is despicted left in green and valid for all dwarfs of the sample. The error bars show the 2-$\sigma$ confidence interval, which corresponds to the errors on the infall mass inference and the 2-$\sigma$ scatter of $z_c$ at a given mass. The shift in the $x$-axis has been arbitrary chosen for the purpose of illustration.}\label{MWUFDHeavy}
\end{figure}

\begin{figure}[h!]
\includegraphics[width=0.48\textwidth]{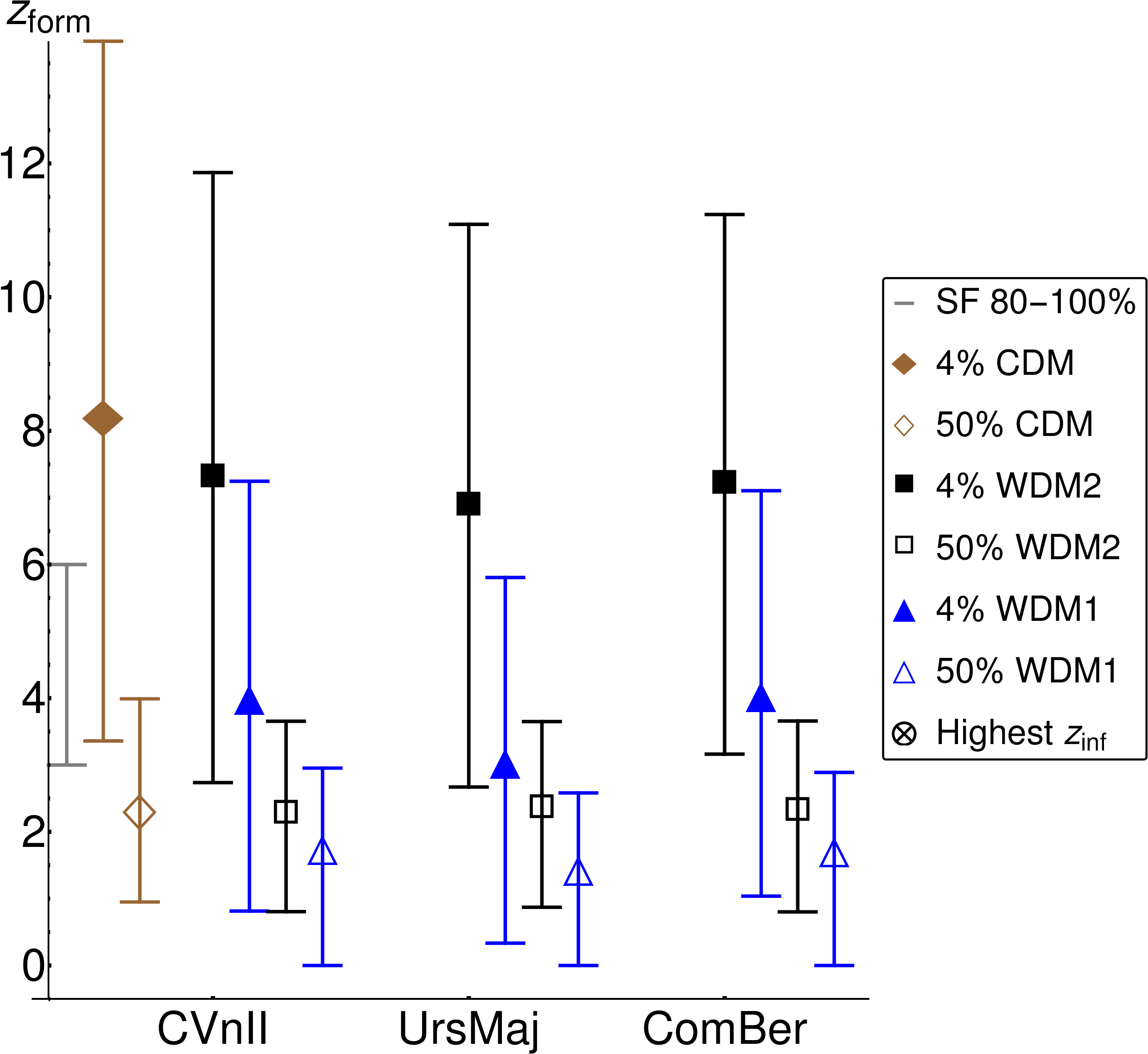}\caption{Collapse redshift distributiont for the three massive UFDs of the MW if their inferred mass would be 10 times heavier ($10^{10}~M_\odot$,, for the different dark matter models indicated in the upper right corner. The redshift by when the galaxies formed 80\% and all of their stars is despicted left in green and valid for all dwarfs of the sample. The error bars show the 2-$\sigma$ confidence interval, which corresponds to the errors on the infall mass inference and the 2-$\sigma$ scatter of $z_c$ at a given mass. The shift in the $x$-axis has been arbitrary chosen for the purpose of illustration.}\label{MWUFDHeavy2}
\end{figure}

For the UFDs, we expect the results of the 50\% criterion to not change since the difference with the time of the onset
of SF, $z \geq 3$, is too large anyway (Figure \ref{zcM}).
On the contrary, if we choose the 4\% criterion, the 2~keV model easily admits now $z>3$ for a subhalo infall mass of $10^{10}~M_\odot$. 
However, considering that the scatter in $z_c$ is large even in the case of halos of 10$^9~M_\odot$, the results of 
\citeauthor{brook14} do not actually weaken much our analysis for the three UFDs with largest inferred infall masses, Ursa Major, Coma Berenice and Canes Venatici II, { see Figure \ref{MWUFDHeavy} that is similar to Figure \ref{summaryMWUFD}. These results do not change our previous conclusions. However this argument only holds for the 2 and 3~keV model. For the 1keV model, the errors that used to be small because of the low collapse value become also important for the three UFDs with large inferred infall masses (here $10^{10}~M_\odot$.), see Figure \ref{MWUFDHeavy2}. The 1~keV model is then not necessarily excluded as strongly, the 4\% criterion has then also large uncertainties. Generally, for heavy halos, all the models will become indistinguishable in the power spectrum, as shown by the 2~keV results which are close to CDM. Heavy halos fit the SFHs more easily and hence make the CDM and WDM scenarios less distinguishable by our method.
\\However we do not expect the UFDs to have such a high mass halo, even before tidal stripping. Such a mass is similar to that of halos of classical dSphs, for which mass modeling is much better established, so it would be surprising that the star-formation efficiency was so much lower in UFDs (which have much less stars). Viceversa, 
tidal stripping could have reduced the mass of both dark matter and stars, but \cite{Tomozeiu} showed that in extreme
stripping events an object with the luminosity of Bootes can be obtained but not
the many other UFDs, like ComBer, that are even fainter.
\\ 
Baryonic effects could potentially weaken the conclusions for the three UFDs with the lowest infall mass $10^8~M_{\odot}$, Hercules, Leo IV and Bootes I. However
for these we have already assumed an infall mass ten times higher than the standard procedure a la Wolf et al. would have suggested, so our conclusions for the light UFDs do not change.

Since there are no high-resolution, self-consistent simulations with baryonic physics currently, the former remains one of the largest unknown in any cosmological simulation and hence being probably responsible for most part of the systematic errors in our model too. Thus it is difficult to estimate an absolute error for the infall mass procedure, while we discussed how some physical processes, such as tidal stirring, dynamical friction and galaxy formation paradigm, rather favor low infall mass.}

The extended Press-Schechter approach assumes the linear perturbation growth, which is thought to be able to reproduce 
the statistical properties of structure formation. The full non-linear regime would, however, add to the scatter between $z_c$ and $M_{infall}$.
One could opt to derive the average growth factor from the \citeauthor{Schechter} theory.
More WDM numerical simulations are needed ultimately. \cite{Maccio10}, using
zoom-in DM-only simulations, concluded that the formation and accretion times for a 2~keV WDM model do not 
differ significantly from CDM, though 
the WDM subhalos form on average slightly later than in CDM. 
 How large is the difference will depend
on which suhalo mass one chooses, though.
Note that in their Figure 3 the difference
in number of subhalos as a function of  formation time between WDM (2 keV) and CDM appears at $z > 9-10$
but this is done for subhalos with infall masses $M(z_{acc}) > 10^9 M_{\odot}$, which
is high compared to what we adopt in our paper for the faintest and most
constraining dwarfs (some of the UFDs). Of course at higher mass scales
it becomes more difficult to discriminate between WDM models at a few keV
and CDM. Their choice to focus on relatively large
subhalos are likely forced by resolution limits in their simulations.
Therefore we believe they do not provide
enough systematic information across a wide range of sunhalo masses at infall
to provide a thorough comparison with our EPS calculations. Pure dark matter simulations
 will not be conclusive as they
are not in the CDM case (and a trivial rescaling of the baryonic effects based on the CDM simulations with hydrodynamics is also potentially
flawed since small halos grow differently and would have not only a different timing for the onset of star-formation but also 
possibly a different amount and pattern of star-formation over time, see e.g. \cite{governato}.)

There are also some uncertainties due to the fact that our method is using an inhomogeneous set of data with different intrinsic errors and
systematics, while ideally, a consistent way of deducing the star-formation should be used for the whole sample.

Finally, there are uncertainties about the nature of the warm particle itself, which ultimately implies the simple-minded notion according to which 
WDM is just CDM with a truncation of the power at some small enough scale is only one of the many possible realizations.
The problem indeed is that bounds from structure formation 
actually constrain the free-streaming length, rather than the ``raw'' mass of the particle. 
Most numerical simulations of structure formation are based on the assumption that warm DM particles are produced via a mechanism 
that gives them a thermal spectrum. If this assumption does not hold anymore, it might become difficult to draw general conclusions. One 
example is the resonant thermal mechanism, which can be seen as a superposition of a WDM component and a non-thermal cold component. 
In the latter case the particle could have a lower rest-mass energy than in the standard thermal relic WDM scenario while 
still being consistent with the Lymann $\alpha$ forest constraints \citep{Drewes}. Only recent numerical simulations have thoroughly studied structure formation with resonantly-produced sterile neutrinos \citep{Bozek}.

\section{Conclusion}\label{conclusion}
We have introduced a new, simple method to { possibly} exclude WDM scenarios via the timing of the onset of star-formation in the oldest dSph and
ultra-faint galaxy satellites in the LG. 
{ Most of the galaxies of our sample, e.g. for which the inferred infall mass does not exceed a few $10^9~M_\odot$, could disfavor at 2-$\sigma$ statistical level a particle of 1~keV, consistently with other results. In the 2 and 3~keV models, large uncertainties weaken our 
results but three UFDs could still disfavor the 2~keV model at that level. With their early SFHs and low-mass, 
they are indeed the objects less affected by our various caveats. }
The simplicity of our method suggests that, with better SF data and more robust theoretical predictions of collapse times in WDM
directly coming from simulations, it can potentially be critical in excluding or admitting WDM models in the critical
range of a few keVs. It offers clues of what to look for in future
numerical hydro-simulations of satellite galaxy formation in the WDM scenario.
Improving the time resolution in the star-formation histories of LG dSphs, as well as increasing the number of UFDs with SF histories
measured at least at the same level of time resolution of the 6 used in this paper,
will allow to increase the constraining power of our analysis by reducing the errors. We note that 
there are other dwarfs, such as Andromeda XIII \citep{weisz},
with strong evidence in favor of  a very early mass assembly but which do not have available kinematics and
therefore were not included our sample. These objects can potentially strengthen the case for excluding
the 2 and 3~keV models.
In passing we note that we have not included
in our samples the distant dSphs Cetus and Tucana as well as the transition dwarfs, such as Phoenix and LGS3, for which accurate
determinations of the SF histories exist (e.g. \citealt{Gallart15}). This is because these are all bright dwarfs that would yield
weak constraints as the classical dSphs as their infall masses will inevitably be predicted to be in the high end of the distribution.

As we discussed, biases, and not just errors might be
hidden behind our determination of the infall mass as presented in this paper.
In order to gain accuracy in the infall mass determination
the analysis could be repeated adopting dark matter density profiles that take into account the effects of baryonic physics,
such as proposed by Brook \& Di Cintio (2014). This however requires that such new profile models be calibrated with
a variety of baryonic feedback recipes rather than with only specific recipes, which will eventually be accomplished.
In the meantime we believe our analysis based on NFW profiles for dark matter subhalos is able to produce a simple proof-of-concept 
description of the new constraining method proposed in this paper.


\acknowledgments{
We thank Aurel Schneider for extensive help on the calculation of the collapse redshift based on extended Press-Schechter theory, and for providing
feedback on early results of our work. We also thank Alexandre Refregier and his group members at ETH Z\"urich for useful comments during 
the development of the Master thesis on which this paper is based. We thank the referee for the useful questions raised that improved the presentation of this work. }

\newpage

\appendix{
\section{Dwarf galaxies properties \label{dwarfspropi}}

Table \ref{dwarfdwarf} lists several
properties of the studied dwarf galaxies such as their stellar mass, velocity dispersion, half-light radius, and dynamical masses. We used the review by \citealt{mc}.

\begin{table}[h!]
\begin{center}
\begin{tabular}{ l  l  l  l  l }
{Galaxy} & $\mathbf{M_{\bigstar}}$ [$10^6~M_{\odot}$] & $\mathbf{\sigma_*}$ [km\,s$^{-1}$]]& $\mathbf{r_h}$ [pc] & $\mathbf{M_{dyn}}$ [$10^6~M_{\odot}$] \\ \hline
Carina & 0.037 & 6.6 & 250 & 6.3\\
Canes Venatici I & 0.23 & 7.6 & 564 & 19 \\
Leo I & 5.5 & 9.2 & 251 & 12 \\
Leo II & 0.74 & 6.6 & 176 & 4.6\\
Sculptor & 2.3 & 9.2 & 283 & 14\\ 
Fornax & 20 & 11.7 & 710 & 56 \\ 
Ursa Minor & 0.29 & 9.5 & 181 & 9.5 \\ 
Draco & 0.29 & 9.1 & 221 & 11 \\ 
 \hline
Bootes I & 0.029 & 2.4 & 242 & 0.81\\ 
Canes Venatici II & 0.0079 & 4.6 & 74 & 0.91\\ 
Coma Berenices & 0.0037 & 4.6 & 77 & 0.27\\ 
Hercules & 0.037 & 3.7 & 330 & 2.6\\ 
Leo IV & 0.019 & 3.3 & 206 & 1.3 \\ 
Ursa Major & 0.014 & 7.6 & 319 & 11 \\
 \hline
Andromeda I & 3.9 & 10.6 & 672 & 44 \\ 
Andromeda II & 7.6 & 7.3 & 36 & 105 \\ 
Andromeda III & 0.83 & 4.7 & 479 & 6.1\\
Andromeda VII & 9.5 & 9.7 & 776 & 42\\ 
Andromeda XI & 0.049 & $\leq$4.6 & 157 & 1.9 \\
Andromeda XII & 0.031 &2.6 & 304 & 1.2\\
\end{tabular}
\caption{Summary of general dwarf galaxy properties, regrouped in our 3 sets: the Milky Way bright classical dSphs, the MW ultra-faint dwarfs and the Andromeda satellites \textit{Column 1}: Galaxy name. \textit{Column 2}: Stellar mass $M_{\bigstar}$ of the dwarf, assuming a light-to-mass ratio of 1. \textit{Column 3}: Observed velocity dispersion $\sigma_*$ of the stellar component. \textit{Column 4}: Half-light radius $r_h$, corresponding to the radius which encloses half of the total density of the stars. \textit{Column 5}: Dynamical mass $M_{dyn}$ of the dwarfs, corresponding to the mass enclosed within the half-light radius following \citealt{walker} with $M_{dyn}=580r_h\sigma_*^2$. \textit{Reference}: \citealt{mc} \label{dwarfdwarf}}
\end{center}
\end{table}

\section{MW bright satellites and Andromeda satellites\label{MW}}
Figure \ref{summaryMW} shows the results for the bright satellites of the Milky Way in the left panel. The galaxies are reported in order of 
increasing halo mass from left to right, namely from Carina at the left corner to Draco at the right corner. 
In the right panel we show the Andromeda satellites for which SFHs have been determined, here reported just in order of numerical suffix from left
to right,
For each galaxy the redshift at which 90\% of stars are formed is plotted in green, and the collapse redshifts, i.e. the redshifts 
at which the halo had accreted 4\% and 50\% of the material, respectively, in the 2 and 3~keV models, are plotted 
in red and blue colors. The redshifts are deduced from the infall mass of each individual galaxy by applying the method explained in \ref{methods} 
and using the curves in Figure \ref{zcM}. 

\begin{figure}[h!]
\includegraphics[width=.49\textwidth]{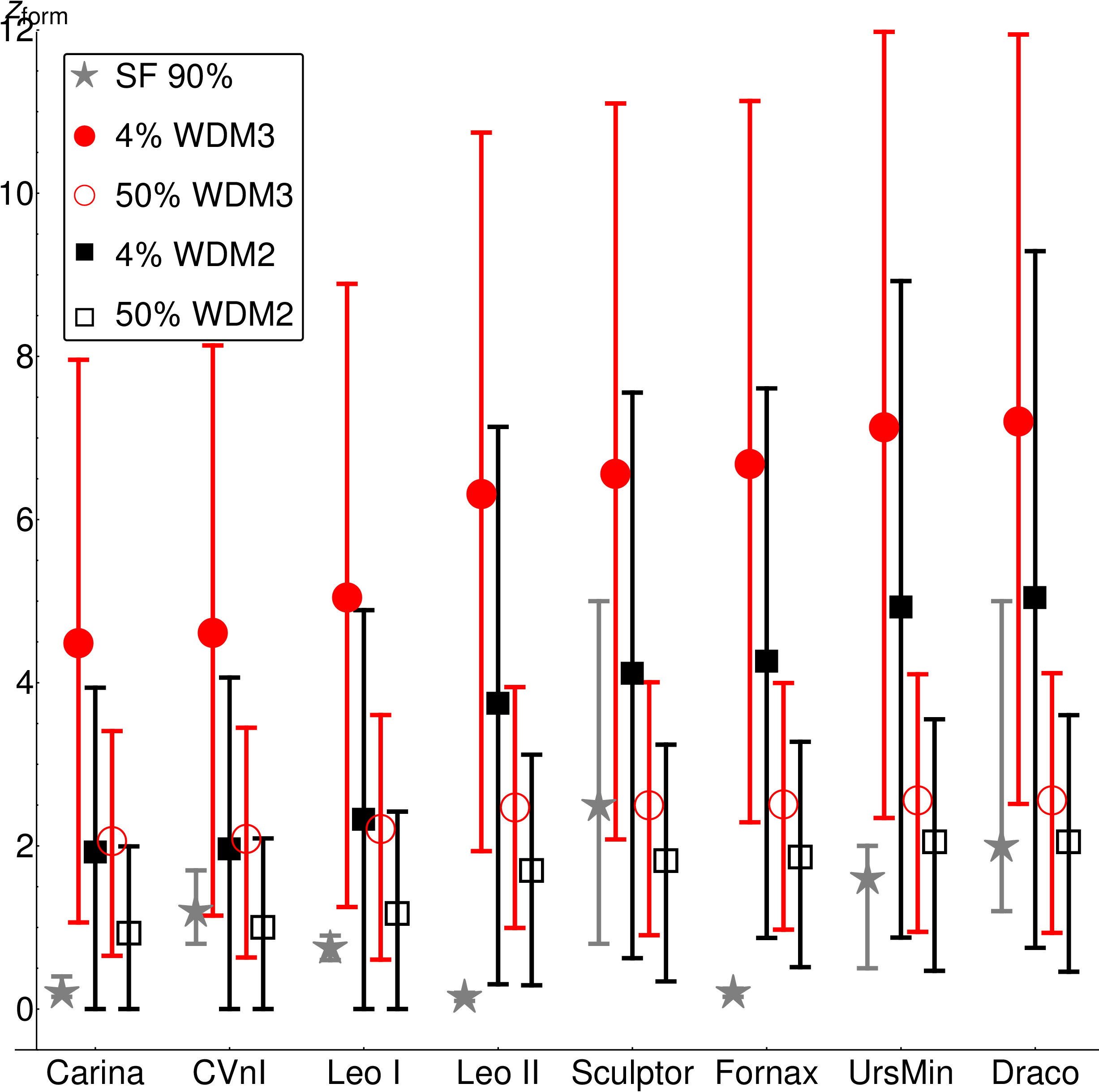}
\hspace{0.5cm}
\includegraphics[width=.47\textwidth]{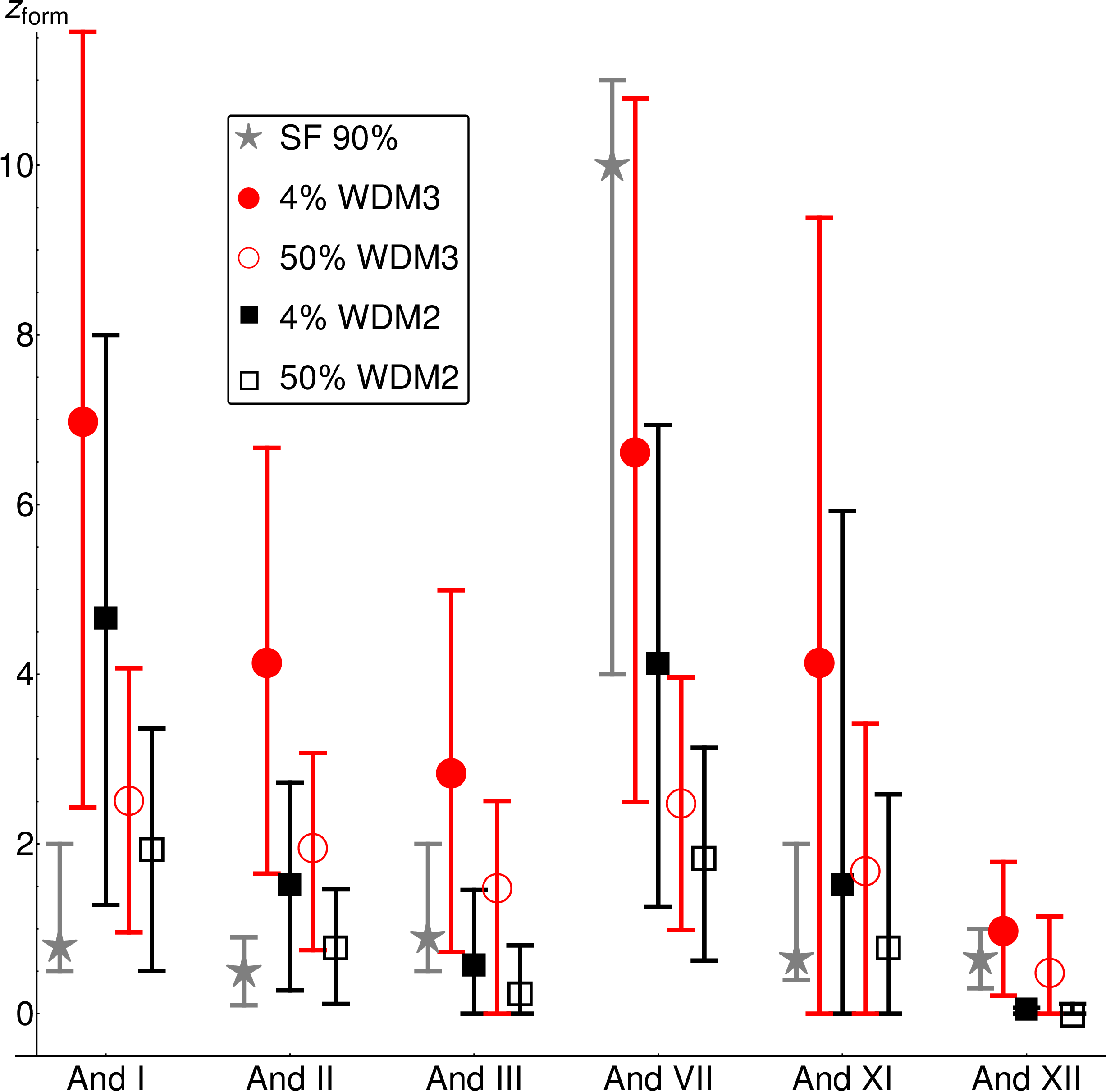}
\caption{Collapse redshift distribution for the Milky Way classical satellites and seven Andromeda satellites, for the different dark matter models indicated in the upper left corner. For each dwarf, we also show the redshift by when the galaxies formed 90\% of their stars, depicted left in green. The shift in the $x$-axis has been arbitrary chosen for the purpose of illustration.
}\label{summaryMW}
\end{figure}
%
We note that, while the constraints from SFHs are not as stringent as for the UFDs described in the main text, Draco, Sculptor and CVnI 
are only marginally consistent with collapse redshifts expected in the 2 and 3~keV models. Clearly more accurate SF histories would be beneficial.
Among Andromeda satellites, And VII, and And XII only for the 3~keV model, pose a stronger constraint since at least the 50\% collapse redshift criterion 
is inconsistent with their SFH. The population of And dwarfs is large and only for a subset we have SF histories, suggesting a potentially significant room for improvement for the predictive power of our analysis.

\section{Dynamical friction\label{DF}}

A dwarf galaxy of mass $M$ orbiting in the Milky Way typically traverses a distribution of stars and dark matter, this latter being the dominant component. The individual stars and the dark matter distribution are deflected by the gravitating mass $M$, hence the density of matter behind the satellite galaxy is greater than in front of it. The wake provokes a gravitational attraction on $M$, which decelerates its motion. As a final consequence, the satellite is slowly falling toward the center of the galaxy host. The phenomenon is called dynamical friction. 

Typically, to reach the center, the satellite needs
\begin{equation}
t_{fric}=\frac{19\mathrm{Gyr}}{\ln\Lambda}\left(\frac{r_i}{5\mathrm{kpc}}\right)^2\frac{\sigma}{200\mathrm{km}\,\mathrm{s}^{-1}}\frac{10^8\mathrm{M}_\odot}{M}. \label{tfric}
\end{equation}
To reach a distance $r_f\neq0$, the satellite needs 
\begin{equation}
t_{fric}=\frac{19}{\ln\Lambda}\left(\frac{r_i^2-r_f^2}{25\mathrm{kpc}}\right)\frac{\sigma}{200\mathrm{km}\,\mathrm{s}^{-1}}\frac{10^8\mathrm{M}_\odot}{M}~[\mathrm{Gyr}]. \label{tfricrf}
\end{equation}
For an initial orbit radius with typical values $r_i\sim100~$kpc, $\sigma\sim200$~km/s for the MW and $M\sim10^9~M_\odot$ for the satellite, we get $t_{fric}\sim$~110 Gyr. For a satellite of that mass, the dynamical friction is not efficient since the mass does not reach the center of the MW in a time comparable to the age of the universe. However with a heavy halo infall mass (see the discussion in the main text \ref{dfric}), the satellite experiences more friction and only needs 17 Gyr to fall into the center. Furthermore, the dispersion velocity of the MW is known to be $\sim200$~km/s at $z=0$ but lower at higher redshifts, typically at $z\sim2$, based on simulations but also on scaling arguments in extended Press-Schechter theory. Indeed the Milky Way had a lower mass at earlier epochs, hence a lower dispersion velocity \cite{}.

In addition to the pure dynamical friction, the MW satellites also experience tidal stripping. As they orbit, they lose mass. As a result, their progression to the center is delayed. \cite{monci} showed that the mass decays exponentially, hence the frictional time is about a factor $e$ higher: 
\begin{equation}
\tau_m=1.2\frac{J_{cir}r_{cir}}{[GM/\textrm{e}]\ln(M_{MW}/M)}\epsilon^{0.4}~[\mathrm{Gyr}],
\end{equation}
where $J_{cir}$ is the angular momentum per unit mass, $r_{cir}$ the radius of the circular orbit, $\epsilon$ an eccentricity factor which is typically $\sim0.7$ for dwarfs thus $\epsilon^{0.4}\sim 1$.
Table \ref{tabledynfric} shows $t_{fric}$ for different $M$, $\sigma$ and $r_f$ and with or without the tidal stripping effect.

We infer from Table \ref{tabledynfric} that only the heavy infall masses in a young Milky Way (so that its velocity dispersion $\sigma$ is small) experience a significant dynamical friction. For the purpose of our analysis, we look at the current distance to the center of the MW of the heaviest dwarfs and use Equation \eqref{tfricrf} to compute the time they needed to orbit from a typical infall distance of 100~km/s to their current orbital distance. Most of the dwarfs are known to be not too close of the MW center (except Sagittarius and Coma Berenices), hence the dynamical friction they experienced so far gives a shorter time than the characteristic values listed in the table above. The timescale for two of the UFDs, Canes Venatici II and Ursa Major, drops even more since their galactic-centric distances is $\sim$100~kpc, or greater. Once we computed the time, we converted it to a redshift via the standard time-redshift relation and the lookback time definition. We take the distances listed in \cite{wolf} and use $\sigma\sim100$~km/s. 

\begin{table} [h!]
\centering
\begin{tabular}{l  l  l  l  l  l}
$\sigma=200$~km/s & $10^9~M_\odot$ & $10^{10}~M_\odot$ & with e & $10^9~M_\odot$ & $10^{10}~M_\odot$ \\ \hline
$t_{fric}\rightarrow$ 0~kpc & 111 Gyr & 17 Gyr & & 329 Gyr & 50 Gyr\\
$t_{fric}\rightarrow$ 50~kpc & 83 Gyr & 12 Gyr & & 247 Gyr & 37 Gyr \\ \hline
$\sigma=100$~km/s & & & & & \\
$t_{fric}\rightarrow$ 0~kpc & 69 Gyr & 12 Gyr & & 164 Gyr & 24 Gyr \\
$t_{fric}\rightarrow$ 50~kpc & 52 Gyr & 9 Gyr & & 123 Gyr & 18 Gyr\\ \hline
$\sigma=50$~km/s & & & & &
 \\
$t_{fric}\rightarrow$ 0~kpc & 47 Gyr & 11 Gyr & & 82 Gyr & 12 Gyr \\
$t_{fric}\rightarrow$ 50~kpc & 52 Gyr & 8 Gyr & & 62 Gyr & 9 Gyr\\ \hline
\end{tabular}
\caption{The frictional time $t_{fric}$ for different $M$, $\sigma$ and $r_f$. The right part of the table displays the merging time with the tidal stripping effect, this is why the values are higher, except for very low $\sigma\sim50~$km/s where these differences are lessened.}\label{tabledynfric}
\end{table}

\section{Comparison of the SFH methods\label{SFH}}
We compare the results of \cite{brown14} with our reference \cite{weisz} when results are available in table \ref{z90UFD}. The values are in both methods in favour of an early SF stopping except for CVnII, though the set of available data is limited.
\begin{table}[h!]
\centering
\begin{tabular}{l|l l|l l}
dSph & \multicolumn{2}{c|}{$z_{SFH=0.9}$(Weisz et al.)} & \multicolumn{2}{c}{$z_{SFH=0.8-1}$(Brown et al.)} \\ \hline \hline
\multirow {2}{20 mm }{CVnII} & \multirow{2}{8 mm}{1.1} & + 0.6 & \multirow {2}{8 mm }{4.5} & + 1.5 \\
& & -- 0.8 & & -- 1.5 \\ \hline
\multirow {2}{20 mm }{Hercules} & \multirow {2}{8 mm }{7.5} & + 0.5 & \multirow {2}{8 mm }{4.5} & + 1.5 \\
& & -- 7.0& & -- 1.5 \\ \hline
\multirow {2}{20 mm }{LeoIV} & \multirow {2}{8 mm }{4} & + 1.0 & \multirow {2}{8 mm }{4.5} & + 1.5 \\
& & -- 3.4 & & -- 1.5 \\
\end{tabular}\caption{Comparison of the two methods of SFHs. The table shows the time by when the galaxy formed 90\% of its stars, with the total uncertainties. \label{z90UFD}}
\end{table}
}

\end{document}